# Acoustic Phonon Characteristics of β-Ga$_2$O$_3$ Single Crystals Investigated with Brillouin-Mandelstam Light Scattering Spectroscopy

Dylan Wright,[1,2,*] Erick Guzman,[1,*] Md. Sabbir Hossen Bijoy,[3] Richard B. Wilson,[4] Dinusha Herath Mudiyanselage,[5] Houqiang Fu,[5] Fariborz Kargar,[3,×] and Alexander A. Balandin[1,2,×]

[1]Department of Materials Science and Engineering, University of California, Los Angeles, California 90095 USA

[2]California NanoSystems Institute, University of California, Los Angeles, California 90095 USA

[3]Materials Research and Education Center, Department of Mechanical Engineering, Auburn University, Auburn, Alabama 36849 USA

[4]Department of Mechanical Engineering, Materials Science and Engineering, University of California, Riverside, California 92521, USA

[5]School of Electrical, Computer, and Energy Engineering, Arizona State University, Tempe, Arizona 85281 USA

[*]Contributed equally to this work
[×]Corresponding authors: fkargar@auburn.edu (F.K.); balandin@seas.ucla.edu (A.A.B)

# ABSTRACT

We report an investigation of the bulk and surface acoustic phonons in β-Ga$_2$O$_3$ ultra-wide bandgap single crystals along various crystallographic directions using Brillouin-Mandelstam spectroscopy. Pronounced anisotropy in the acoustic phonon dispersion and velocities was observed across different crystal orientations. The measured average acoustic phonon velocities for the crystallographic directions of interest are $\bar{v}_{(001)} = 5{,}250$ ms$^{-1}$ and $\bar{v}_{(\bar{2}01)} = 4{,}990$ ms$^{-1}$. The surface acoustic phonons propagate approximately twice as slowly as the bulk acoustic phonons. Our results suggest that the anisotropy of heat conduction in β-Ga$_2$O$_3$ results from the difference in phonon velocities rather than the phonon lifetime. The obtained information for bulk and surface acoustic phonons can be used for developing accurate theoretical models of phonon scattering and optimization of thermal and electrical transport in this technologically important ultra-wide bandgap semiconductor.

Ultra-wide bandgap (UWBG) semiconductors have drawn widespread interest in recent years for their potential applications in power electronics, UV optoelectronics, and RF electronics[1–7]. UWBG semiconductors are a class of semiconductors, which includes diamond, AlN, BN, and $Ga_2O_3$ with bandgaps higher than 3.4 eV, the band gap of GaN. Among the UWBGs, $Ga_2O_3$, particularly its monoclinic β-phase, has attracted attention due to its large bandgap of ~4.4-5.0 eV and a breakdown field as high as ~8 MV/cm² [8–17]. However, despite their high breakdown field, $Ga_2O_3$ devices experience substantial Joule heating under high power densities[18,19]. The latter explains the need for accurate knowledge of $Ga_2O_3$'s thermal properties for reliable modeling of heat dissipation[20]. Recent research into thermal conductivity in this material system has revealed the values of ~10-20 Wm⁻¹K⁻¹ [21–23] which is an order of magnitude lower than that in GaN with thermal conductivity of ~130-230 Wm⁻¹K⁻¹ [24–29]. Acoustic phonons are the leading heat carriers in these semiconductors. The significant difference in thermal conductivity of GaN and $Ga_2O_3$ comes despite comparable longitudinal (LA) and transverse (TA) acoustic phonon group velocities[30,31]. The volumetric heat capacity of $Ga_2O_3$ is only 20% lower than that of GaN[32,33], implying that the low thermal conductivity of $Ga_2O_3$ is primarily due to the short lifetime of its acoustic phonons, *i.e.*, strong scattering. Prior studies have also demonstrated a strong anisotropy of heat conduction[22,32,34]. The above considerations motivated our experimental study, which focused on the direct measurement of acoustic phonon frequencies and velocities, as well as the extraction of phonon lifetimes from the spectral peak width. The bulk and surface acoustic phonon characteristics are also needed for the theoretical description of electron–phonon scattering, as well as for understanding polaron and polariton effects in such a class of materials[35–37].

For this study, we used (001) and ($\bar{2}$01) β-Ga₂O₃ single crystals obtained by edge-defined film-fed growth (Novel Crystal Technology, Inc)[38]. Both samples had low full width at half maximum (FWHM) of 20-30 arcsec from their X-ray diffraction rocking curves, indicating good crystallinity and low dislocation density. The two samples had similar thickness of ~650 μm and n-type Sn dopant concentration on the order of ~5×10¹⁸ cm⁻³. The sample with (001) surface orientation was prepared with an offcut angle of 0° in both the [100] and [010] in-plane directions, and the sample with ($\bar{2}$01) surface orientation was prepared with an offcut angle of −0.1° in the [010] and −0.6° in the [102] in-plane directions. β-Ga₂O₃ adopts a monoclinic crystal structure with lattice parameters $a = 12.214$ Å, $b = 3.0371$ Å, $c = 5.7981$ Å, and interaxial angles $\alpha = \gamma = 90°, \beta = 103.83°$ [39].

Raman and Brillouin-Mandelstam light (BMS) scattering measurements were conducted on both samples to investigate their optical and acoustic phonon properties and elucidate the anisotropic characteristics of phonons.

Raman spectroscopy measurements (Renishaw inVia Qontor) were performed using 488 nm and 633 nm excitation lasers in a backscattering configuration. Before data collection, the instrument was carefully calibrated using a Si reference sample to ensure the phonon peak appeared at 520 cm$^{-1}$ for both excitation wavelengths. In the Raman measurements, the polarization of the incident light is selected to be parallel to the $b$ crystallographic direction for the (001) sample and parallel to the [102] direction in the ($\bar{2}$01) sample. The polarization of the scattered light was not analyzed. To minimize potential laser-induced heating, the laser power was kept below 2 mW in all experiments. Note that, in the Raman backscattering configuration, the magnitude of the probed phonon vector is defined as $|q| = [n_i \omega_i + n_s \omega_s]/c$, where $c$, $\omega$, and $n$ are the speed of light in vacuum, angular frequency, and index of refraction, respectively. Subscripts $i$ and $s$ represent the incident and scattered light[40]. As the laser excitation wavelength changes, the $q$ varies from 21.52 μm$^{-1}$ at 633 nm to 30.38 μm$^{-1}$ at 488 nm, allowing one to probe phonon energies as a function of their wave vector, and thereby obtaining dispersion. Given the backscattering geometry used in all Raman experiments, the direction of the probed phonon wavevector is determined by the surface orientation. For the (001) sample, $q$ aligns with $\boldsymbol{G_{(001)}} = \boldsymbol{c^*}$, while for the ($\bar{2}$01) surface, it aligns with $\boldsymbol{G_{(\bar{2}01)}} = -2\boldsymbol{a^*} + \boldsymbol{c^*}$, where $\boldsymbol{a^*}$, $\boldsymbol{b^*}$, and $\boldsymbol{c^*}$ are the reciprocal lattice vectors of the monoclinic structure.

β-Ga₂O₃ has a monoclinic crystal structure, with a primitive cell containing 10 atoms, resulting in 3 acoustic and 27 optical phonon branches. The irreducible representation of these modes at the Γ point is $\Gamma = 10A_g + 5B_g + 4A_u + 8B_u$ where the $A_g$ and $B_g$ symmetries are Raman-active and can be observed depending on the polarization and scattering geometry[41]. Figure 1 presents the Raman spectra of β-Ga₂O₃ measured with light incident perpendicular to the (001) plane using both laser excitation wavelengths. The inset shows the crystal structure along with the (001) plane. The spectra were fitted using individual Lorentzian functions, which are represented by colored peaks in Figure 1. All ten $A_g$ modes were detected in both cases. Note that in this scattering configuration, the $B_g$ modes are expected to be Raman inactive. However, we note that some of these Raman-silent modes, indicated by arrows, were detected, likely due to the relaxation of selection rules caused by the presence of Sn dopants in the structure.

Similar experiments were conducted on ($\bar{2}$01) β-Ga₂O₃ using the 488 and 633 nm excitation lasers. Figure 2 shows the Raman spectra for β-Ga₂O₃ probed perpendicular to the ($\bar{2}$01) plane. All modes with $A_g$ symmetries were detected on this sample, except the one located at 764 cm$^{-1}$. The spectral position and FWHM of the observed peaks for both samples are listed in Table 1. As seen, for each sample, the peak spectral positions, *i.e*, the phonon energies, shift with the excitation wavelength, reflecting the characteristics of optical phonon dispersions near the Brillouin zone center. The spectral positions of all modes shift within ~3 cm$^{-1}$. We calculated the approximate group velocity of optical phonons as $v_g \sim (\Omega_b - \Omega_r)/(q_b - q_r)$, where $\Omega$ is the angular frequency of the phonon band observed in Raman, $q$ is wavevector, and subscripts *b* and *r* correspond to Raman results obtained by 488 nm blue and 633 nm red laser excitations, respectively. The calculated $v_g$ of all modes falls below $100 \text{ ms}^{-1}$, within the calculational error, showing that all modes have flat dispersion in the vicinity of the BZ center. The FWHM values of these modes are nearly identical for both samples, within the instrumental error, indicating similar phonon decay rates for each crystal orientation.

We now turn to the results of Brillouin–Mandelstam spectroscopy (BMS), also referred to as Brillouin-light scattering (BLS), of the *β*-Ga₂O₃ single crystals. BMS is a nondestructive optical technique that probes zone-center low-energy acoustic phonons in the GHz frequency range[42]. Due to its distinct optical configuration and the need to prevent high-intensity reflected light from entering the spectrometer, BMS measurements are generally conducted with laser light incident on samples at oblique angles in the backscattering configuration. Changing the incident angle with respect to the normal to the sample has another unique advantage. It enables one to tune the direction of the phonon wavevector. As a result, one can obtain the energy of the acoustic phonons as a function of phonon wavevector and estimate the anisotropic phonon group velocities in different crystallographic directions. All BMS experiments in this study were conducted in the backscattering configuration using a 532 nm laser excitation wavelength. The details of our BMS procedures are reported elsewhere for other material systems [43–49].

The real-space lattice vectors *a*, *b*, and *c* in the monoclinic structure of *β*-Ga₂O₃, are not mutually orthogonal. To simplify the analysis of the scattering experiments, we introduce a *fixed* Cartesian coordinate system, referred to as the lab-frame '*xyz*' system. This coordinate system, illustrated schematically in Figure 3 (a-b), is defined such that the *z*-axis is aligned with the sample surface normal, N, of the (001) and ($\bar{2}$01) *β*-Ga₂O₃ samples. The scattering plane, formed by the incident and scattered wavevectors, lies in the *xz*-plane of the lab frame for the (001) sample and in the *yz*-plane for the ($\bar{2}$01) sample. In this configuration, the crystal orientation relative to the fixed laboratory axes differs for each surface, allowing access to different phonon propagation directions while keeping the lab-frame scattering geometry constant, the incident light is *p*-polarized for all measurements. Under this configuration, the angle between the electric field vector of the incident light and the *b*-axis is $90° − \theta$. As *β*-Ga₂O₃ is optically birefringent, deviation of the incident polarization from the *b*-axis at oblique angles can result in beam splitting. Each resulting ray within the medium may contribute to scattering processes with bulk phonons within the crystal. The phonon wavevector, *q*, lies along the direction of the diffracted rays, which deviate from the surface normal by an angle $\theta^*$ (Figure 3 (c)). This angle is calculated using Snell's law as $\theta^* = \sin^{-1}\{\sin(\theta)/n_{eff}\}$ where $n_{eff}$ is the effective refractive index experienced by the diffracted ray and can be determined based on $n_j$, the refractive indices along principal axes. The orientation of the crystal with respect to the lab-frame scattering plane is further defined by the azimuthal angle $\zeta$, which is the in-plane rotation angle between the *b*-axis and the *y*-axis in the (001) sample, as shown in Figure 3(d). This angle is defined as the angle between [102] and the *x*-axis for the (201) sample.

The directional indices $n_j$ were determined through ellipsometry and calculated from the Sellmeier equation $n_j^2(\lambda, T) = \varepsilon_{1,j}(T) + A \cdot \lambda^2/(\lambda^2 - \lambda_i^2)$, where $\varepsilon_{1,j}$ is the temperature-dependent dielectric permittivity for the direction $j$, and $A = 0.57$ and $\lambda_i = 0.27$ μm are the Sellmeier coefficients specific for β-Ga$_2$O$_3$[22]. From these models, we obtained refractive indices for the doped β-Ga$_2$O$_3$ sample as $n_{[100]} = 1.91$, $n_{[010]} = 1.96$, and $n_{[001]} = 1.94$. Our measured values are $n_{[100]} = 1.89$ and $n_{[010]} = 1.915$, ~1% and ~2.5% lower for light polarized in the [100] and [010] directions in the (001) sample. In the ($\bar{2}$01) sample, we observe an identical value for the [010] direction and $n_{[102]} = 1.96$. The magnitude of $\boldsymbol{q}$ is derived from momentum conservation as $|q| = (4\pi n_{eff}/\lambda_{BMS})$.

Figure 4 (a) shows the BMS spectra of (001) β-Ga$_2$O$_3$ as a function of $\theta$, corresponding to the scattering geometry shown in Figures 3 (a, b). The angle of incidence, $\theta$, was varied from 20° to 70°, while the in-plane azimuthal angle was maintained at $\zeta = 0°$. In BMS, typically, three peaks appear on each side of the spectra, corresponding to the longitudinal acoustic (LA) and two transverse acoustic (TA$_1$, TA$_2$) phonons. In our results, between $\theta = 20°$ and 30°, only a single TA peak is observed, indicating that the two TA modes are degenerate. At small $\theta$, the phonon wavevector $q$ aligns closely with the $\boldsymbol{c^*}$ reciprocal lattice vector, thus probing phonons along the $\Gamma - Z$ direction. Prior DFT calculations of phonon band structure show that the TA modes are degenerate near the Brillouin zone (BZ) center along this high symmetry direction[50], in agreement with our observations. However, as $\theta$ increases, $q$ deviates from the $\Gamma - Z$ direction, lifting the degeneracy of TA phonons due to the low-symmetry monoclinic structure of β-Ga$_2$O$_3$. As the incident angle measurements progress from 20° to 70°, the frequency, $f$, of the LA mode decreases from 56.2 to 54.3 GHz, while that for TA$_1$ increases from 27.4 to 32.1 GHz. The frequency of the lower branch TA$_2$ phonon mode remains almost unchanged.

Figure 4 (b) shows the results of BMS experiments performed at a fixed incident angle $\theta = 60°$ while varying the azimuthal angle, $\zeta$. Changing $\zeta$ effectively rotates the direction of the probing phonon wavevector around a cone with a deviation angle of $\theta^*$ relative to the surface normal. This experimental setup enables the investigation of how acoustic phonons propagate along different crystal directions and planes, revealing the material's strongly anisotropic acoustic properties. A notable feature is the emergence of a shoulder on the LA peak starting at $\zeta = 30°$, which we attribute to the scattering of LA phonons by birefringence-induced ray splitting. This effect arises from the optical anisotropy of the crystal, where the same LA phonon branch is probed via two slightly different phonon wavevectors, each corresponding to a distinct effective refractive index. The same type of shoulders can be seen in peaks associated with TA phonons. The evolution of the spectral position of these modes as a function of $\zeta$ is presented in Figure 4 (c). As $\zeta$ increases, the energy of TA$_1$ phonon mode decreases, reaching a minimum at $\zeta = 45°$, and then increases as the sample completes a 180° rotation. In contrast, the TA$_2$ branch exhibit the opposite trend, with its energy increasing up to $\zeta = 80°$, followed by a decrease as the rotation progresses. There is an intriguing crossing of the TA phonon branches at $\zeta = 120°$. The LA phonon branch shows minimal variation with $\zeta$, indicating weak angular dependence.

Similar BMS experiments were performed on $(\bar{2}01)$ β-Ga$_2$O$_3$ as a function of incident angle, $\theta$, and azimuthal angle, $\zeta$, shown in Figures 5 (a, b), respectively. For incident angle dependence, the scattering plane is normal to $(\bar{2}01)$ surface, parallel to the [102] direction, with the incident light being $p$-polarized. At small $\theta$, the phonon wavevector is close to $\boldsymbol{G} = -2\boldsymbol{a}^* + \boldsymbol{c}^*$ reciprocal lattice direction. As $\theta$ increases, $q$ deviates from this direction, accompanied by a decrease in the LA phonon energy, an increase in the TA$_1$ mode energy, and a nearly constant TA$_2$ mode. Figure 5 (b) shows the dependence of the LA and TA phonons spectral positions as a function of $\zeta$ at a fixed $\theta = 60°$. The phonon energies exhibit a symmetric trend with respect to $\zeta = 90°$. This symmetry is more comprehensively visualized in Figure 5 (c), which presents the full evolution of frequency as a function of $\zeta$ for all observed modes.

Our BMS measurements on (001) and $(\bar{2}01)$ samples reveal pronounced anisotropy in the acoustic phonon branches along different crystallographic directions. Given that acoustic phonons contribute significantly to heat transport in β-Ga$_2$O$_3$ at room temperature[51], it is crucial to assess how this anisotropy affects thermal conduction. Previous studies have shown that the cross-plane thermal conductivity of (001)-oriented β-Ga$_2$O$_3$ is slightly higher than that for $(\bar{2}01)$-oriented sample. In crystalline solids, thermal conductivity can be expressed as $k = (1/3)C\bar{v}\Lambda$, where $C$ is the volumetric heat capacity, $\bar{v}$ is the average phonon group velocity, and $\Lambda$ is the average phonon mean free path. Since $\Lambda = \bar{v}\tau$, where $\tau$ is the average phonon lifetime, and thermal conductivity simplifies to $k = (1/3)C\bar{v}^2\tau$. The average phonon lifetime can be expressed as $\tau = 1/\eta$, where $\eta$ is the phonon scattering rate.

Acoustic phonons typically possess higher group velocities than optical phonons. Consequently, the average phonon velocity $\bar{v}$ relevant to thermal transport can be estimated from the group velocities of acoustic modes extracted from the spectral positions of the BMS peaks. Knowing the frequency of the phonon modes from BMS measurements and the corresponding phonon wave-vector, $q$, one can obtain the phase velocity, $v_P$, of the acoustic phonons as $v_p = 2\pi f/q$. Near the Brillouin zone (BZ) center, the dispersion of fundamental LA and TA polarization branches is linear, *i.e.*, $\omega = qv_p$, which implies that the phase velocity, $v_p = \omega/q$, is equivalent to the group velocity, $v_g = \partial\omega/\partial q$, *i.e.*, $v_p = v_g$. To minimize uncertainties arising from anisotropic refractive indices in the calculation of the phonon wavevector, we analyzed spectra acquired from both (001) and ($\bar{2}$01) at a fixed incident angle of $\theta = 20°$ and azimuthal angles of $\zeta = 0°$. In this geometry, the polarization of the incident light is nearly aligned with the *b*- and [102]- crystallographic directions for the (001) and ($\bar{2}$01) samples. We use our measured refractive indices of $n_{[010]} = 1.915$ and $n_{[102]} = 1.96$ to calculate the phonon wavevector for each sample. For the (001) sample, the measured spectral positions are $f_{LA} = 56.2$ GHz and $f_{TA_{1,2}} = 24.1$ GHz yielding group velocities of $v_{LA} = 7,810$ ms$^{-1}$ and $v_{TA} = 3350$ ms$^{-1}$. For the ($\bar{2}$01) sample, the spectral positions are $f_{LA} = 52.1$ GHz, $f_{TA_1} = 27.4$ GHz, and $f_{TA_2} = 23.8$ GHz resulting in $v_{LA} = 7,070$ ms$^{-1}$, and $v_{TA_1} = 3,720$ ms$^{-1}$ and $v_{TA_2} = 3,230$ ms$^{-1}$. The root mean square average of the acoustic phonon velocities is calculated using $\bar{v} = \sqrt{(1/3)\{v_{LA}^2 + v_{TA_1}^2 + v_{TA_2}^2\}}$ yielding $\bar{v}_{(001)} = 5,250$ ms$^{-1}$ and $\bar{v}_{(\bar{2}01)} = 4,990$ ms$^{-1}$.

The thermal conductivity ratio between the two orientations is expected to scale with the square of the average group velocity, i.e., $\bar{v}_{(001)}^2/\bar{v}_{(\bar{2}01)}^2$. Using the obtained results from BMS experiments, we obtain the ratio of $\bar{v}_{(001)}^2/\bar{v}_{(\bar{2}01)}^2 \sim 1.11$, indicating that the cross-plane thermal conductivity for the (001)-oriented β-Ga$_2$O$_3$ is expected to be approximately 11% higher than the ($\bar{2}$01)-oriented samples. Previous reports indicate that the thermal conductivity of β-Ga$_2$O$_3$ for the (001) and ($\bar{2}$01)-oriented β-Ga$_2$O$_3$ is 13.7 and 11.4 Wm$^{-1}$K$^{-1}$ [32], respectively, showing a ~20% difference, which is close to the ratios calculated from BMS experiments. The long-wavelength, small $q$, acoustic phonons, probed by BMS, do not contribute strongly to the phonon Umklapp scattering. The phonon lifetime, which is more relevant to thermal resistance, can be estimated from the FWHM of Raman-active optical phonon modes. Our measurements indicate that the FWHM of the optical phonon modes is nearly identical for both crystallographic orientations, suggesting that the phonon lifetimes, and thus scattering rates, are comparable. This observation is consistent with prior theoretical calculations[21]. Therefore, the anisotropy in thermal conductivity is attributed primarily to differences in phonon group velocities along different crystallographic directions.

In conclusion, we investigated the bulk and surface acoustic phonons in UWBG β-Ga$_2$O$_3$ single crystals using Brillouin-Mandelstam spectroscopy. The anisotropy in the acoustic phonon velocities agrees well with the available thermal conductivity data. The obtained information for bulk and surface acoustic phonons can be used to develop accurate theoretical models of phonon scattering and optimize thermal and electrical transport in this UWBG semiconductor.

**Acknowledgments**

The work at UCLA and ASU was supported by ULTRA, an Energy Frontier Research Center (EFRC) funded by the U.S. Department of Energy, Office of Science, Basic Energy Sciences under Award # DE-SC0021230. A.A.B. and F.K. acknowledge the support of the National Science Foundation (NSF) via a Major Research Instrument (MRI) DMR Project No. 2019056 entitled "Development of a Cryogenic Integrated Micro-Raman-Brillouin-Mandelstam Spectrometer."

**Conflict of Interest**

The authors declare no conflict of interest.

**Author Contributions**

F.K. and A.A.B. conceived the idea, coordinated the project, and led the data analysis and manuscript preparation. E.G. performed the Brillouin and Raman spectroscopy and contributed to data analysis. D.W. performed Brillouin spectroscopy and conducted data analysis. R.W. conducted the ellipsometry measurements and assisted with data analysis. M.B.H.S. contributed to Raman data analysis; D.H.M. and H.F. provided the samples and contributed to the data interpretation. All authors reviewed and contributed to the final manuscript.

**The Data Availability Statement**

The data in support of the findings of this study are available from the corresponding author upon reasonable request.

Acoustic Phonon Characteristics of β-Ga$_2$O$_3$ Single Crystals Investigated with Brillouin-Mandelstam Light Scattering Spectroscopy – 2025

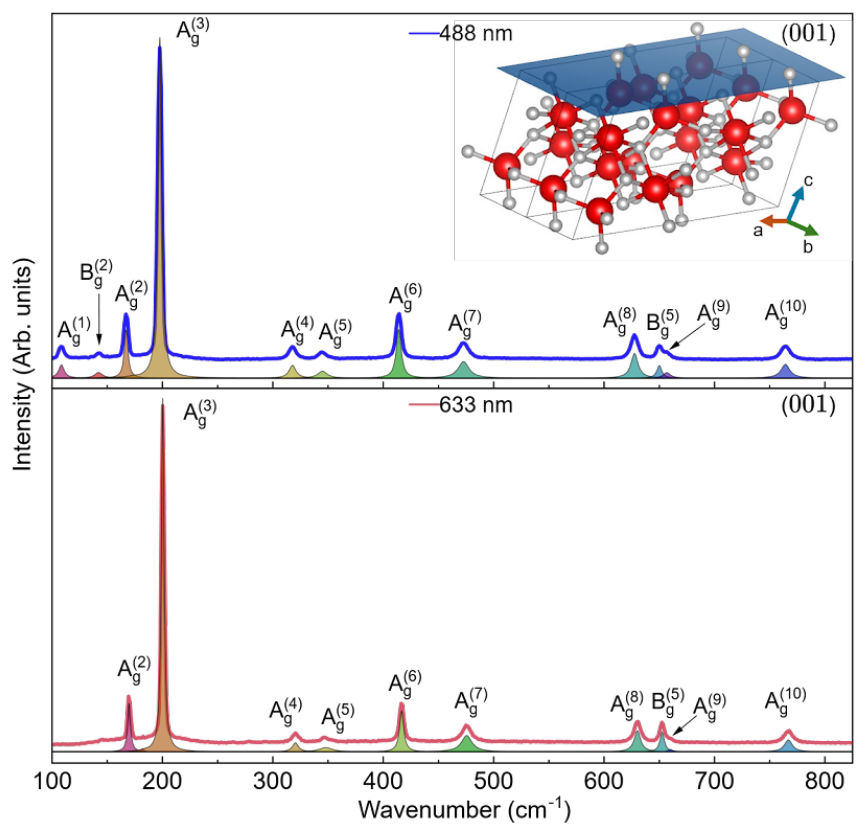

**[Figure 1.** Raman spectra of (001)-oriented β-Ga$_2$O$_3$ measured using 488 nm and 633 nm laser excitation wavelengths. The inset illustrates the (001) plane of the monoclinic β-Ga$_2$O$_3$ crystal structure. Colored peaks represent individual Lorentzian fits applied to the measured data.]

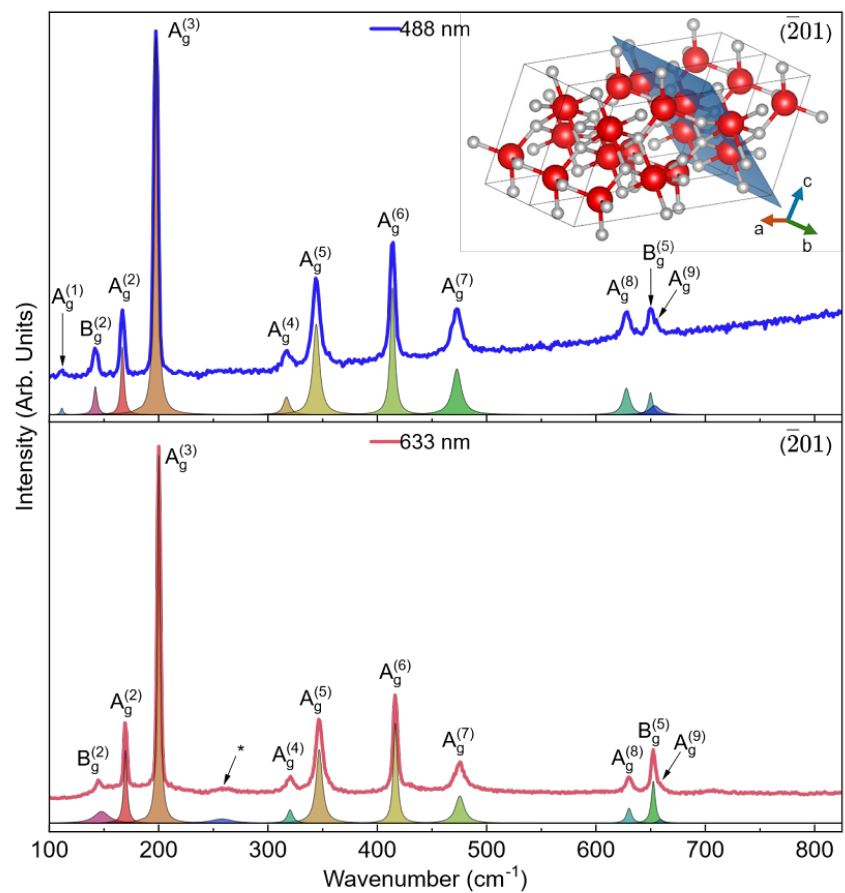

**[Figure 2.** Raman spectra of ($\bar{2}$01)-oriented β-Ga₂O₃ measured using 488 nm and 633 nm laser excitation wavelengths. The inset illustrates the (001) plane of the monoclinic β-Ga₂O₃ crystal structure. Colored peaks represent individual Lorentzian fits applied to the measured data.]

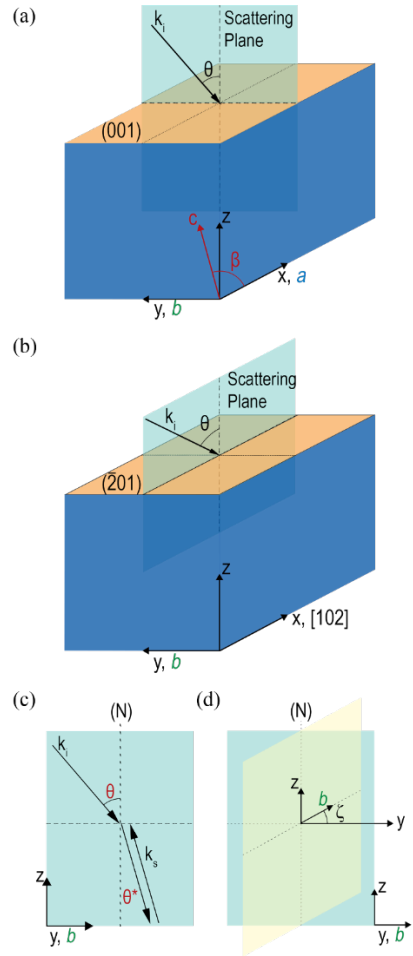

[**Figure 3.** (a) Experiment geometry for the (001) sample. The incident beam vector $k_i$ enters the surface at an incidence angle $\theta$ within the scattering plane (cyan) which lies in the *y-z* plane. The lab-frame *z*-axis is normal to the surface, while the crystal axes *a*, *b*, and *c* are indicated. The monoclinic angle $\beta$ between *a* and *c* is shown in red. (b) Experiment geometry for the ($\bar{2}$01) sample. The scattering plane in this sample lies in the *x-z* plane. (c) Projection of the scattering geometry for the (001) sample in the (100) plane, which serves as the reference plane. The incident ($k_i$) and scattered ($k_s$) wavevectors are shown, forming angles $\theta$ and $\theta^*$ with respect to the surface. (d) The projection of the crystal lattice vectors onto the *y-z* plane for the (001) sample, showing the azimuthal angle $\zeta$ between the crystal *b*-axis and the laboratory *y*-axis. This angle defines the in-plane rotation of the crystal relative to the scattering plane. Note that for the ($\bar{2}$01) sample the reference plane is the (010) plane and $\zeta$ is the angle between the [102] crystallographic direction and laboratory *x*-axis.]

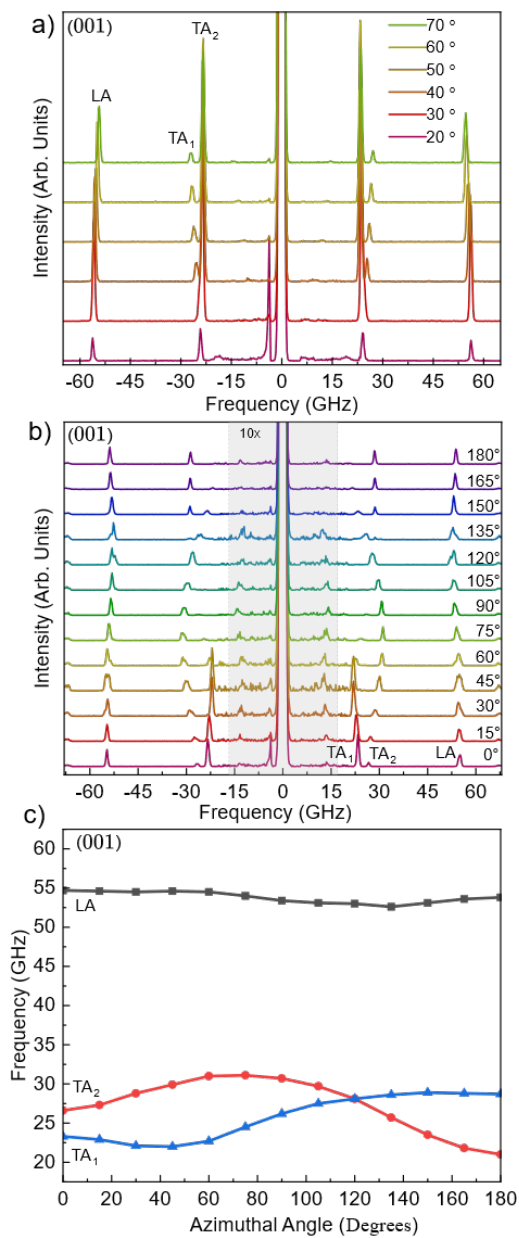

[**Figure 4.** BMS spectra of (001)-oriented β-Ga$_2$O$_3$ as a function of (a) incident angle, $\theta$, and (b) azimuthal angle, $\zeta$. The peaks labeled as LA and TA are associated with longitudinal and transverse acoustic phonons. (c) spectral evolution of LA and TA phonon modes as a function of azimuthal angle, $\zeta$.]

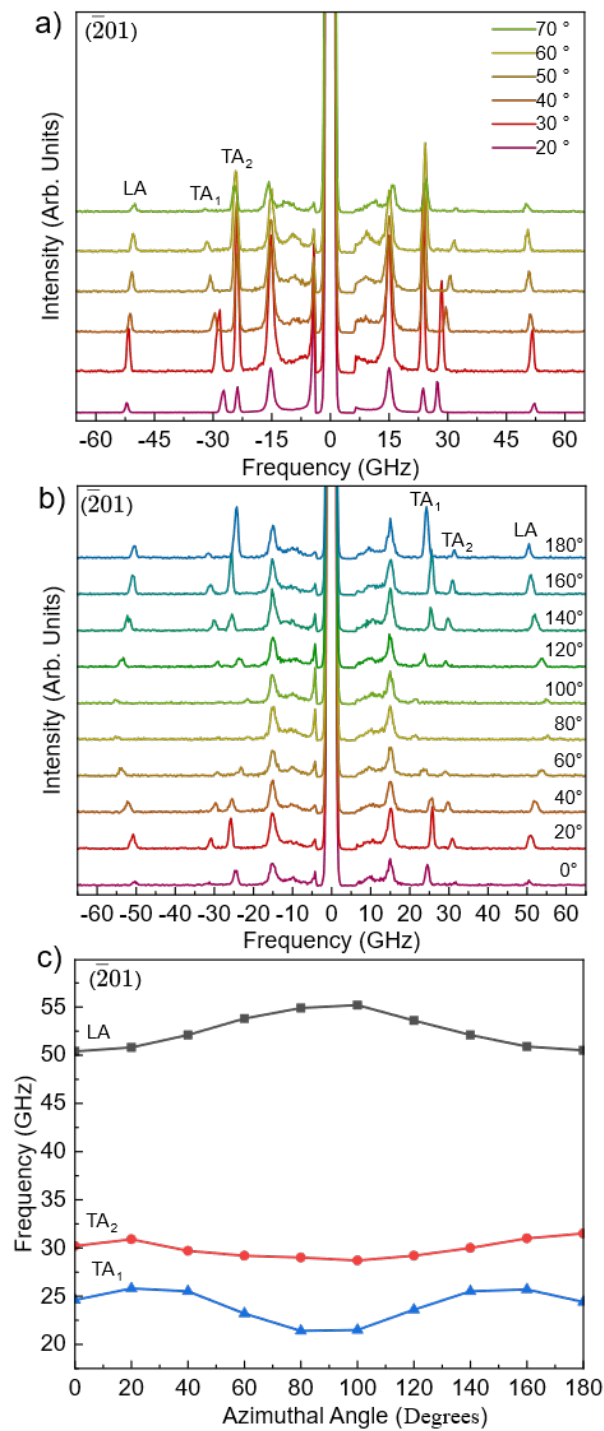

[**Figure 5.** BMS spectra of ($\bar{2}$01)-oriented $β$-Ga$_2$O$_3$ as a function of (a) incident angle, $θ$, and (b) azimuthal angle, $ζ$. (c) spectral evolution of LA and TA phonon modes as a function of azimuthal angle, $ζ$.]

## Table I: Frequency and FWHM of Raman peaks of β-Ga$_2$O$_3$

|  | Wavenumber (cm$^{-1}$) | | | | FWHM (cm$^{-1}$) | | | |
|---|---|---|---|---|---|---|---|---|
|  | (001) | | (-201) | | (001) | | (-201) | |
|  | 488 nm | 633 nm | 488 nm | 633 nm | 488 nm | 633 nm | 488 nm | 633 nm |
| A$_g^{(1)}$ | 108.5 | - | 111.5 | - | 6.1 | - | 2.7 | - |
| B$_g^{(2)}$ | 142.2 | - | 142.1 | 147.8 | 8.3 | - | 3.9 | 16.3 |
| A$_g^{(2)}$ | 167.1 | 169.7 | 166.9 | 169.6 | 4.0 | 3.2 | 3.5 | 3.6 |
| A$_g^{(3)}$ | 197.6 | 200.3 | 197.6 | 200.2 | 3.9 | 3.0 | 3.9 | 3.1 |
| * | - | - | - | 257.8 | - | - | - | 24.1 |
| A$_g^{(4)}$ | 317.8 | 320.4 | 316.9 | 320.2 | 7.4 | 6.6 | 6.8 | 8.1 |
| A$_g^{(5)}$ | 344.9 | 347.9 | 344.1 | 346.9 | 10.0 | 14.5 | 7.2 | 7.4 |
| A$_g^{(6)}$ | 414.0 | 416.6 | 413.9 | 416.5 | 5.4 | 4.6 | 4.9 | 4.3 |
| A$_g^{(7)}$ | 472.7 | 475.5 | 472.8 | 475.5 | 11.7 | 11.5 | 9.9 | 9.8 |
| A$_g^{(8)}$ | 627.6 | 630.1 | 627.8 | 630.3 | 7.7 | 6.9 | 7.7 | 5.7 |
| B$_g^{(5)}$ | 650.0 | 652.6 | 649.9 | 652.4 | 5.4 | 4.7 | 4.6 | 4.4 |
| A$_g^{(9)}$ | 956.9 | 659.9 | 653.7 | 658.9 | 7.6 | 5.2 | 11.7 | 5.5 |
| A$_g^{(10)}$ | 764.2 | 766.9 | - | - | 9.6 | 8.5 | - | - |